\documentclass[prd,showpacs,twocolumn]{revtex4}
\usepackage{graphicx}

\newcommand{\beq}{\begin{equation}}
\newcommand{\eeq}{\end{equation}}
\newcommand{\beqar}{\begin{eqnarray}}
\newcommand{\eeqar}{\end{eqnarray}}

\ifx\undefined\fiverm
     \font\fiverm=cmr5
\fi

\input{prepictex}
\input{pictex}
\input{postpictex}

\begin{document}
\author{Tarun Biswas}
\title{Spiral Galaxy Rotation Curves Without Dark Matter or MOND -- Two Conjectures}
\email{biswast@newpaltz.edu}
\affiliation{State University of New York at New Paltz, \\ New Paltz,  NY 12561, USA.}
\date{\today}
\begin{abstract}
	Usual explanations of spiral galaxy rotation curves assume circular orbits of
	stars. The consequences of giving up this assumption are investigated here.
	In particular,  hyperbolic stellar trajectories are found to be interesting. The two suggested models for
	the production of such trajectories will also explain the observed flat rotation curves without
	the postulation of dark matter or MOND. It is suggested that spiral galaxies may have started as compact
	objects with significant angular momenta and then disintegrated. The first model
	conjectures the existence of a spinning hot disk around a spherical galactic core. The disk is held together by
	local gravity and electromagnetic scattering forces. However, it disintegrates at the edge producing fragments that
	form stars. Once separated from the disk, the stars experience only the centrally directed gravitational force due to
	the massive core and remaining disk. A numerical simulation
	shows that a high enough angular velocity of the disk produces 	hyperbolic stellar trajectories that agree with
	the observed rotation curves. The second model conjectures a significant initial thermonuclear event that produces
	a dust plume along with large
	stars. This dust plume is made of ordinary matter. 	However, it acts like the postulated dark matter in producing
	initial circular trajectories. Unlike dark matter, the plume can be shown to escape the galaxy rapidly causing the
	star trajectories to evolve to  hyperbolic shapes. This process can be seen to produce the observed rotation
	curves due to the initial circular orbits. Also, as the plume dissipates rapidly it does not obfuscate the stars from
	view. Both models have weaknesses as do the currently known models using dark matter or MOND.
	
\end{abstract}
\maketitle
\section{Introduction}
For some time, measured tangential velocities\footnote{The terms ``tangential'' and ``radial'' are used here in reference to
	the galactic center.} of stars in the spiral arms of spiral galaxies have been a challenge to
explain from theory\cite{rubin0,rubin1,rubin2,bosma}. Currently, the most popular model for explaining these velocities
postulates the presence of the so-called dark matter\cite{persic0,persic1,corbelli,gentile,merritt,yegorova,duffy}.
MOND (Modified Newtonian Dynamics) is another possible explanation\cite{mcgaugh0,mcgaugh1,mcgaugh2}. Neither dark matter
nor MOND has been observed directly yet. So, it might be useful to have other
possible explanations -- particularly, if they do not require the postulation of any new kind of matter or laws of
physics. Here, I shall present a couple of such possibilities.

Anyone who has played with firework spinners as a child (or an adult) must have noticed their resemblance to spiral
galaxies. Anyone who has not can always search ``firework spinner'' on YouTube to see it. The trajectories of the glowing
embers in a spinner look like the stars of the spiral arms. Quite obviously, the actual motion of the
stars cannot be observed directly.
But the still picture that we see suggests non-circular trajectories (like the firework spinner) rather than the circular
trajectories as assumed by most analyses. So, here I shall consider the consequences of assuming hyperbolic trajectories.
With such an assumption, one has to answer two major questions:
\begin{itemize}
	\item Do the stars of the spiral arms eventually escape the galaxy?
	\item Why do the stars conspire to have almost the same tangential speed beyond a certain distance from the core?
\end{itemize}
The answer to the first question is ``Yes''. According to this model the ``spiral'' state of a galaxy is just one transient state
in its time evolution. Due to their hyperbolic trajectories, the stars of the spiral arms are expected to escape the galaxy eventually.
The answer to the second question is to be found in the following detailed discussion that includes a numerical simulation.

\section{The Spinner Model}
In the first model, a spiral galaxy is considered to start as a compact spherical core surrounded by a functionally rigid
spinning disk held together by gravity as well as electromagnetic scattering forces (see section~\ref{secdisk}).
Fragments of the disk break off at the edge in the form of stars.
Here, in the simplest version of this model, we will assume that the stars separate from the disk edge with initial velocities
equal to that of the edge. Then we can assume that the disk angular speed remains constant while its radius decreases due to
loss of material in the form of stars. Hence, we conclude that stars separating earlier have greater initial velocities
than stars that separate later.
Once a star separates, it experiences no local forces. Then, the only force on it is the much weaker long range gravitational force
due to the core and the remaining disk.
Hence, stars start off with significant tangential speeds due to the spinning disk and no radial speeds. But, once separated,
they develop nonzero radial speeds and their tangential speeds decrease. The stars that separate later start with smaller
tangential speeds due to the shrinking of the disk. Hence, if the disk shrinks at a certain rate, it could make the
early-separated stars move at roughly the same speed as the later-separated ones. Due to outward radial speeds developed
after separation from the disk, the stars are expected to have hyperbolic trajectories.

As an individual star moves away from the core, its outward radial speed increases and could eventually become measurable.
However, several mitigating factors are expected to make such measurement difficult. First, the density of stars
decreases with increasing radial distance. Second, the stars at greater radial distances are expected to be colder and
dimmer. Third, the radial velocity component must have a component along the line of sight of the observer from Earth
to allow measurement using the Doppler effect.
This would require the star to have the bright galactic core region in its background as seen from Earth. Such a bright
background might wash out the light of the star.

\section{A Numerical Simulation of the Spinner Model}
Let us assume that the disk has constant areal density and it disintegrates at a constant rate to produce stars of equal mass.
So, the area of the disk will reduce at some constant rate $q$. Hence,
\beq
A = A_0 - qt,
\eeq
where $A$ is the area of the disk at time $t$ and $A_0$ the initial area. If $R$ is the radius of the disk at time $t$
and $R_0$ the initial radius, then
\beq
R^2 = R_0^2 - at, \;\;\mbox{where} \;\; a=q/\pi.
\eeq
Hence, the radius of the disk at time $t$ is given by
\beq
R = \sqrt{R_0^2-at}. \label{eqdiskr}
\eeq
It is to be expected that, below a certain minimum radius $R_m$, there will not be enough centrifugal force to produce
more stars.

Now, let there be a total of $N$ stars created by the disk at equal time intervals of $T$. Let the
$i^{\mbox{\tiny th}}$ star have a radial coordinate $r_i$ after it is created. At the time of creation, each star has an
initial radial coordinate equal to the current radius $R$ of the disk given by equation ~\ref{eqdiskr}. The initial radial velocity
is zero. The initial angular momentum is important to record as it is expected to be conserved under the radially directed
gravitational force from the galactic core and disk. If $\Omega$ is the constant angular velocity of the disk, then the initial angular
momentum is $mR^2\Omega$ where $m$ is the mass of the star. This will be different for different stars as they are created at 
different times with different values of $R$. However, for each star this angular momentum will be conserved. As the
trajectory of a star is independent of its mass, the relevant conserved quantity related to angular momentum is
\beq
l_i = r_i^2\dot{\phi}_i = R^2\Omega, \label{eqli}
\eeq
where $\phi_i$ is the angular coordinate of the $i^{\mbox{\tiny th}}$ star,  $\dot{\phi}_i=d\phi_i/dt$ and $R$ is the initial
radial coordinate. So, the tangential velocity of the $i^{\mbox{\tiny th}}$ star at any time is
\beq
v_{ti} = r_i\dot{\phi}_i = \frac{l_i}{r_i}. \label{eqvti}
\eeq

After creation, the $i^{\mbox{\tiny th}}$ star follows the following differential equation.
\beq
\ddot{r}_i-\frac{l_i^2}{r_i^3}+\frac{GM}{r_i^2}+f_c(r_i,R)=0, \label{eqnewt}
\eeq
where $G$ is the gravitational constant, $M$ is the mass of the galactic core and $f_c(r,R)$ is the gravitation acceleration
produced at a distance $r$ from the center by the disk when its radius is $R$. It can be seen that
\beq
f_c(r,R) = G\sigma \int_0^{2\pi} \int_0^R \frac{\rho(r-\rho\cos\theta)d\rho\, d\theta}{(\rho^2+r^2-2\rho r\cos\theta)^{3/2}},
\eeq
where $\sigma$ is the areal density of the disk. The above integral needs to be computed numerically at each stage of the
computation.

\begin{figure*}
 	\includegraphics[height=12cm,width=\textwidth]{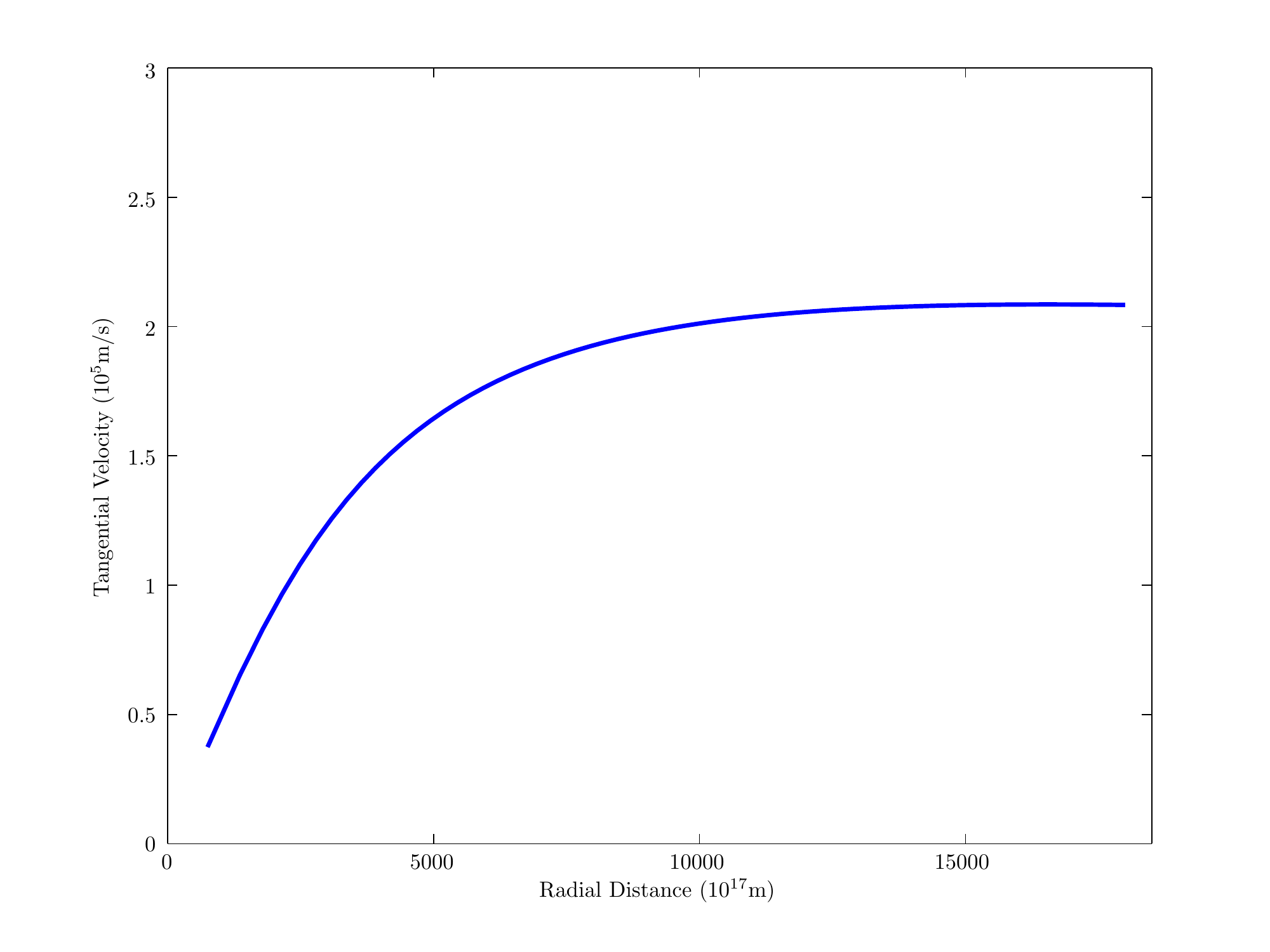}
 	\caption{Rotation curve for $M=10^{35}$kg, $R_0=5.0\times 10^{18}$m, $\Omega=1.5\times 10^{-11}$s$^{-1}$,
 		$\sigma = 0$ and $a=2.0\times 10^{24}$m$^2/$s. \label{fig1}}
 \end{figure*}

\begin{figure*}
 	\includegraphics[height=12cm,width=\textwidth]{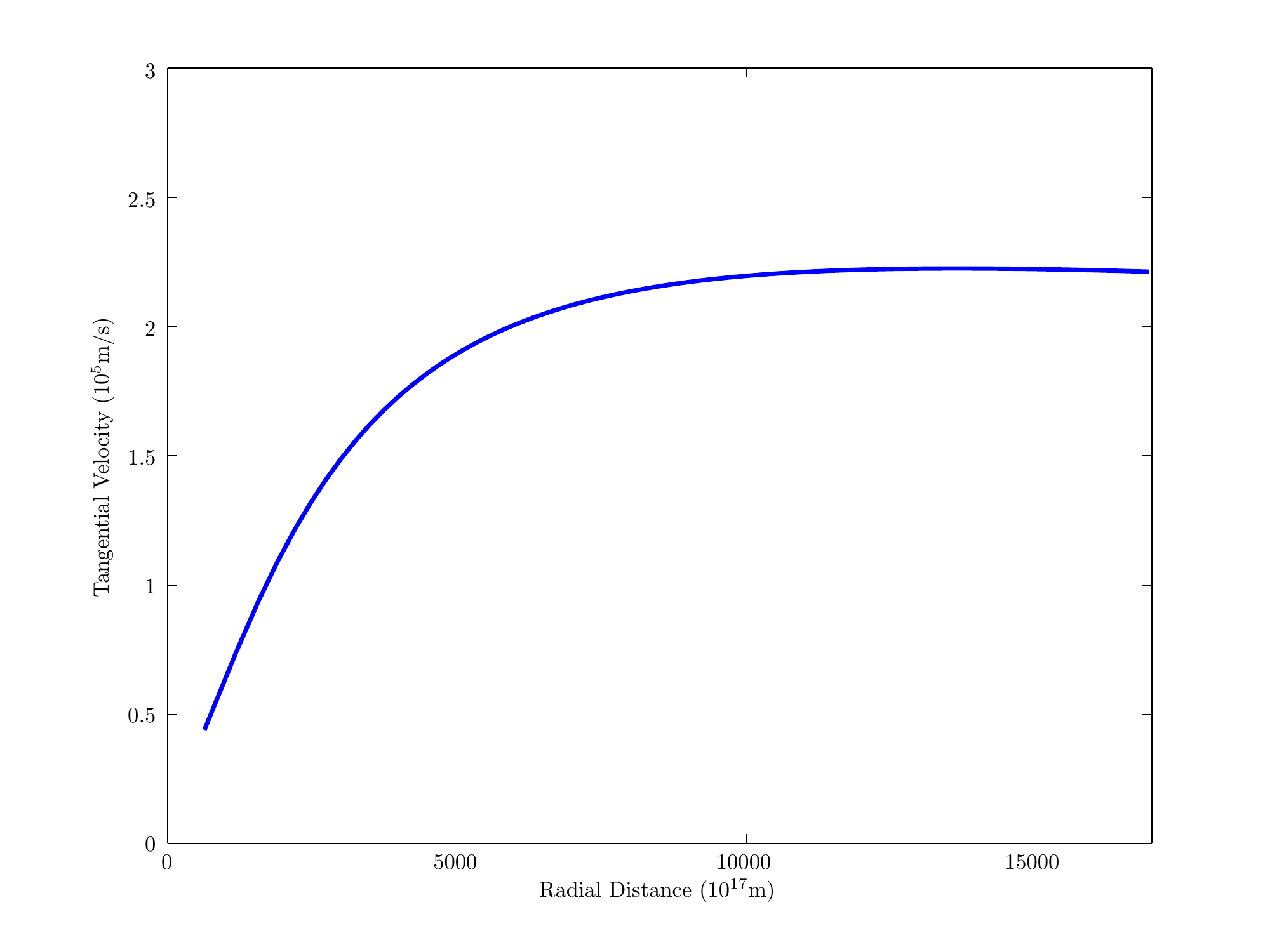}
 	\caption{Rotation curve for $M=10^{40}$kg, $R_0=5.0\times 10^{18}$m, $\Omega=1.5\times 10^{-11}$s$^{-1}$,
 		$\sigma = 0$ and
 		$a=2.0\times 10^{24}$m$^2/$s. \label{fig2}}
 \end{figure*}

\begin{figure*}
 	\includegraphics[height=12cm,width=\textwidth]{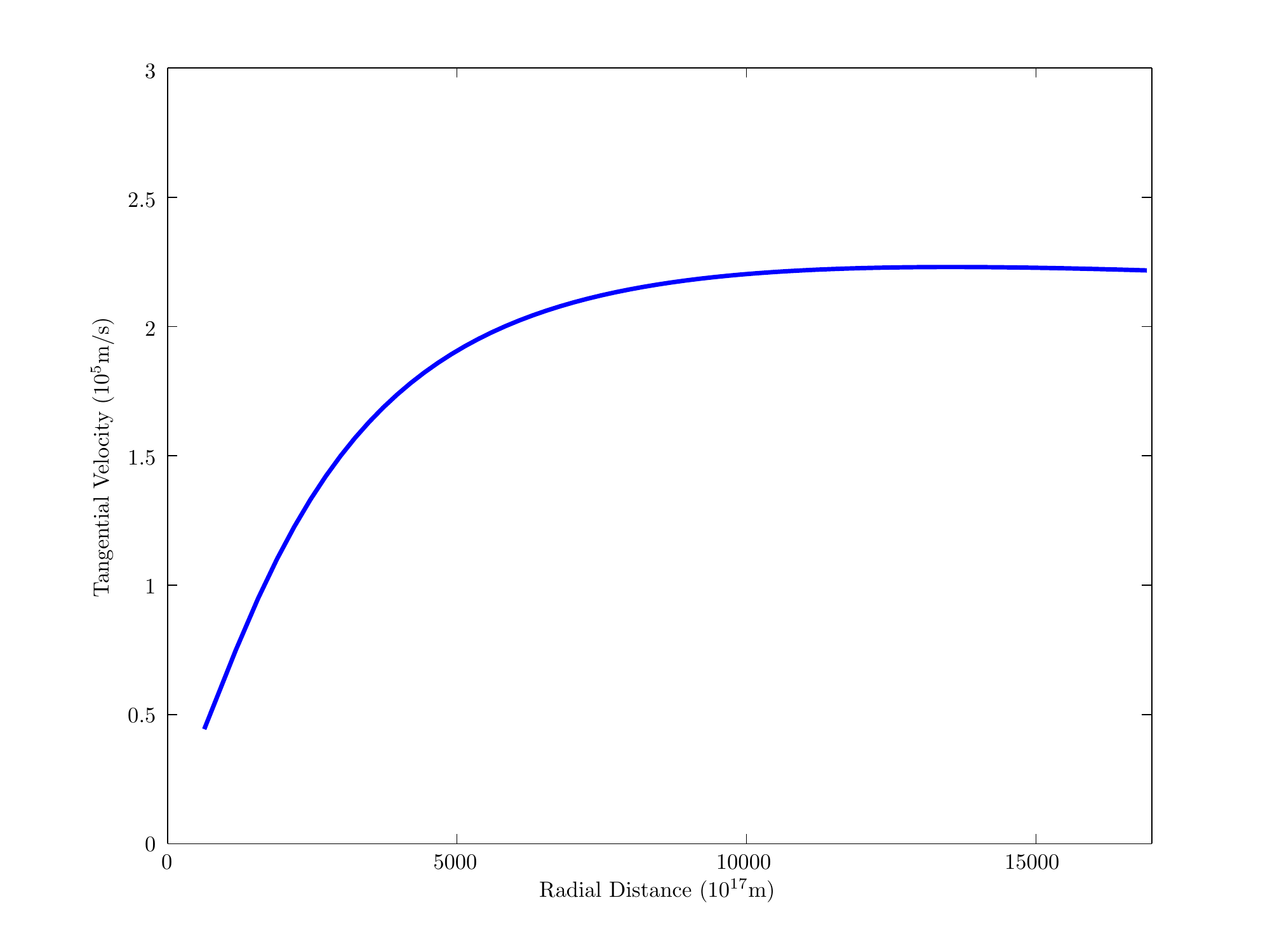}
 	\caption{Rotation curve for $M=10^{40}$kg, $R_0=5.0\times 10^{18}$m, $\Omega=1.5\times 10^{-11}$s$^{-1}$,
 		$\sigma = 10^4$kg$/$m$^2$ and
 		$a=2.0\times 10^{24}$m$^2/$s. \label{fig3}}
 \end{figure*}

\begin{figure*}
 	\includegraphics[height=12cm,width=\textwidth]{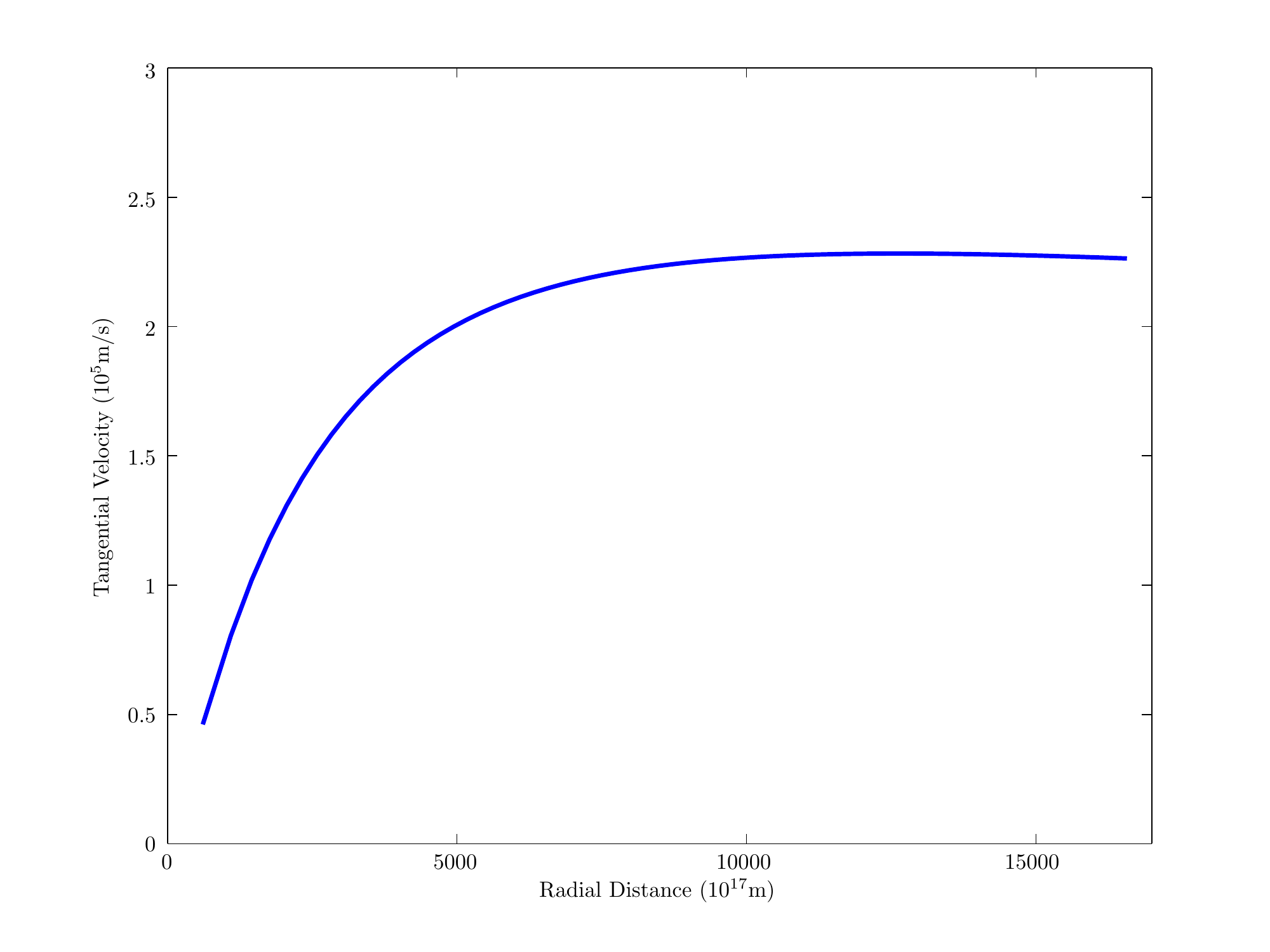}
 	\caption{Rotation curve for $M=10^{40}$kg, $R_0=5.0\times 10^{18}$m, $\Omega=1.5\times 10^{-11}$s$^{-1}$,
 		$\sigma = 10^5$kg$/$m$^2$ and
 		$a=2.0\times 10^{24}$m$^2/$s. \label{fig4}}
 \end{figure*}

\begin{figure*}
 	\includegraphics[height=12cm,width=\textwidth]{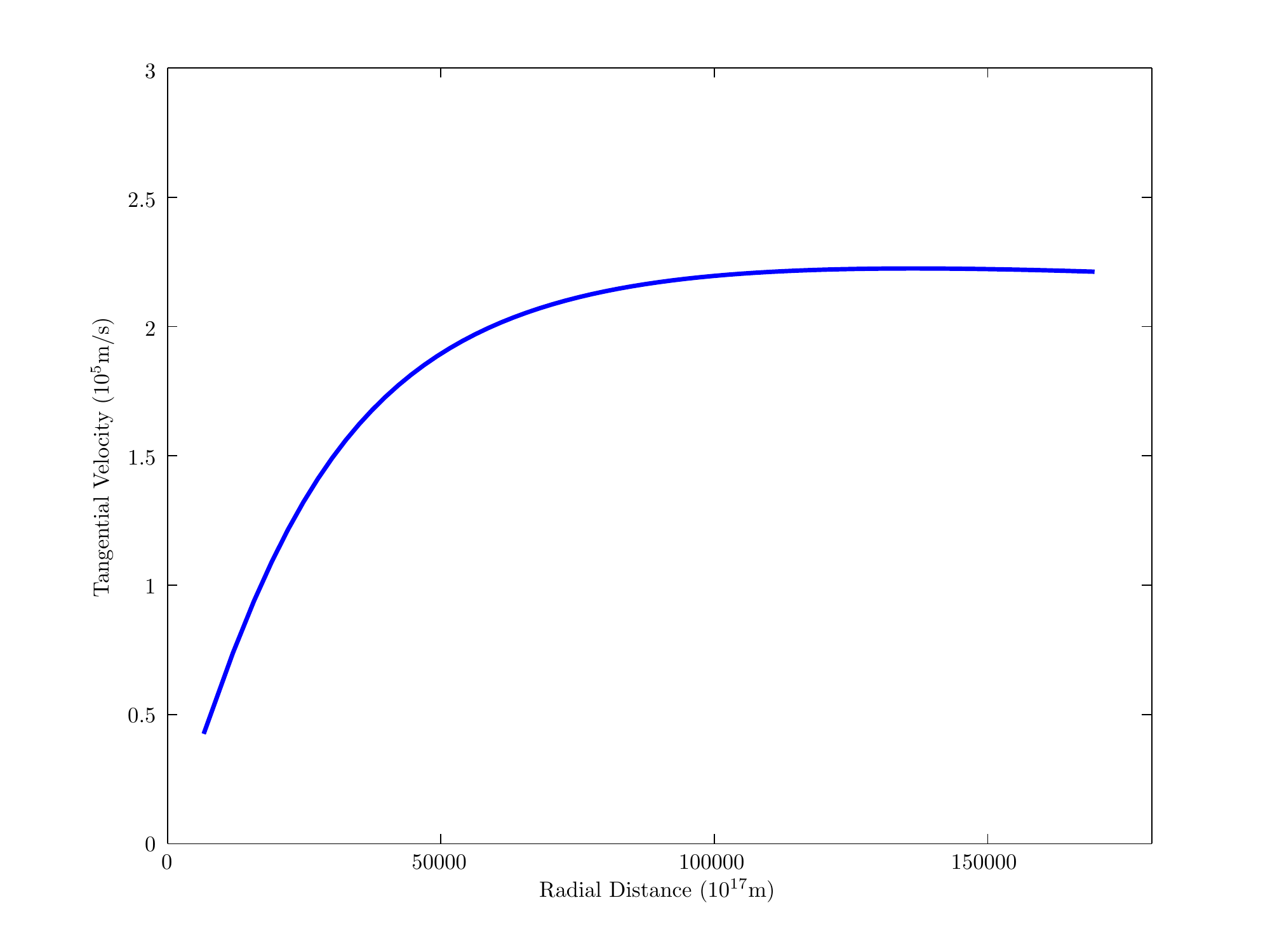}
 	\caption{Rotation curve for $M=10^{40}$kg, $R_0=5.0\times 10^{18}$m, $\Omega=1.5\times 10^{-10}$s$^{-1}$,
 		$\sigma = 0$ and
 		$a=2.0\times 10^{24}$m$^2/$s. \label{fig5}}
 \end{figure*}

The numerical implementation of this simulation is done by looping through the following steps at small intervals of time
$\Delta t = h$ for a total time duration of $NT+T_a$. As stated earlier, $N$ is the number of stars created, $T$ the
time interval at which they are created and $T_a$ is the time elapsed after the last star is created.
\begin{itemize}
	\item If current time $t=iT$ for $i=0,1,2,\dots$, create a new star as long as the disk radius $R$ is greater than
	the minimum radius $R_m$. Find initial $r_i$ using equation~\ref{eqdiskr}.
	Set initial $\dot{r}_i$ to be zero. Find the constant of motion $l_i$ using equation~\ref{eqli}.
	\item Loop through all stars created so far computing their next values for $r_i$ and $\dot{r}_i$ after the time interval
	$h$ using equation~\ref{eqnewt}. Use a fourth order Runge-Kutta algorithm for this purpose.
\end{itemize}
Figures~\ref{fig1} through~\ref{fig5} show some results of this computation. Figures~\ref{fig1} and~\ref{fig2} show that the core
mass does not affect the curves very much. A change of core mass from $10^{35}$kg to $10^{40}$kg produces only a small effect
on the curves. Similarly, figures~\ref{fig2}, \ref{fig3} and~\ref{fig4} show that the disk mass density changing from zero to
$10^4$kg$/$m$^2$ and then to $10^5$kg$/$m$^2$ does not change the curve very much either. This is expected, as most of the speed
of a star comes from its initial speed due to the angular velocity of the disk. This is verified by comparing figures~\ref{fig2}
and~\ref{fig5}. A change in value of $\Omega$ by a factor of 10 changes the curve significantly. After the last star is ejected
from the disk, the first star reaches a much farther point if $\Omega$ is larger.

\section{The Dynamics of the Disk}
\label{secdisk}
In the above analysis, the disk has been assumed to be rigid with constant areal density. This is a first approximation.
In reality, the disk is expected to be a dense collection of massive objects of varying masses. For convenience, we shall call
all such objects ``particles''. The density of the disk is high enough for particles to interact with each other due to gravity
as well as electromagnetic scattering forces in the local neighborhood. So, the forces on individual particles will
not be simply the centrally directed gravitational attraction of the core. The local forces will keep each particle moving
in a roughly circular path as is the case for particles constituting a rigid body. This is the justification for the rigid
body approximation for the disk. Once a star separates from the disk at its edge, the local forces disappear and the long
range centrally directed gravitational forces due to the remaining disk and the core dominate.

\section{The Elvis Model}
For the second model, one may consider the early stage of a spiral galaxy to be a compact ball of matter with
significant angular momentum. This compact
object could explode due to maybe a thermonuclear event. The explosion fragments are expected to be of a wide range of sizes.
Granular particles or even particles of atomic size would be the primary fragments. Let us call this collection of small
fragments the {\bf dust plume}. Inside the plume we would also expect to have some larger objects like stars. Besides
the plume and the stars, a dense and heavy galaxy core is also expected to remain. The gravity of the plume and the core
could keep the stars orbiting in almost circular orbits and produce the observed rotation curve as is explained
by the theory of dark matter. However, unlike dark matter, the plume is expected to block light and hence completely
obfuscate the stars. So, the key argument in this model considers the large range of  speeds of the explosion fragments immediately
after the explosion. Due to partial thermal equilibrium, different fragments are expected to have roughly the same kinetic
energies. If $m$ is the mass of a plume particle and $M$ the mass of a star, then,
\beq
mc^2\left(\frac{1}{\sqrt{1-v^2/c^2}}-1\right)\approx MV^2/2,
\eeq
where $v$ is the speed of the plume particle, $V$ the speed of the star and $c$ the speed of light. Note that we have used
the relativistic formula for the kinetic energy of the plume particle, but not for the star. This is because the plume particle
is expected to reach relativistic speeds but not the star. Hence,
\beq
v\approx c\sqrt{1-\frac{4m^2c^4}{(MV^2+2mc^2)^2}}. \label{eqvel}
\eeq
In the following we shall see that this formula gives $v$ to be practically equal to $c$ and orders of magnitude larger
than $V$. Hence, the plume particles will escape the galaxy rapidly after the explosion leaving behind the stars\footnote{This is
	somewhat like helium atoms escaping from the Earth atmosphere.}.
Once the plume disappears, the stars become visible. However, their positions and velocities do not change very much from the values
determined in the presence of the plume. This is because, for large mass objects, the rates of change of position and velocity
are expected to be small. Hence, the observed tangential velocities of the stars are as they would be in
the presence of the plume which is the same as explained by dark matter. However, once the plume has left, the orbit
shapes cannot be expected to be circular anymore. They are definitely going to be hyperbolic.

So, a spiral galaxy, as we see it now, is a snapshot of the object after the explosion and immediately after the plume has left.
The stars of the galaxy are still visible while they are close to their circular orbit positions and velocities
set immediately after the explosion due to the presence of the plume. The hyperbolic nature of the orbits after the plume
has left, will take some time to become apparent. However, by the time the hyperbolic nature
becomes apparent, the stars will also have moved a large distance away from the galaxy core. At large distances, the density
of stars is expected to be low. As individual stars of a galaxy are not visible, a high enough density of stars is necessary
for the spiral arms to be visible. Hence, the stars farther out in their hyperbolic trajectories, are not visible. This explains
the observed star velocities in the spiral arms.

This model also raises the possibility that the different types of galaxies are all the same but at different stages of their
evolution. In the presence of the plume, the larger stars are not visible.

For an order of magnitude calculation, consider a plume particle of mass $m=1\times 10^{-3}$kg and a star of mass
$M=1\times 10^{30}$kg. Let the speed of the star be $V=1\times 10^5$m/s. Then, from equation~\ref{eqvel}, we
get,
\beq
v\approx c.
\eeq
If the size of the galaxy is $R=1\times 10^{20}$m, then the time of escape for the plume particle is
\beq
T_e\approx R/c \approx 3\times 10^{11}\mbox{ sec}.
\eeq
In the same time a star moves a distance
\beq
d\approx VT_e \approx 3\times 10^{16}\mbox{ m}.
\eeq
$d$ being negligibly small compared to $R$, the stars seem to stay practically stationary while the plume escapes. Hence, Doppler shift
is the only means of measuring their speed.

\section{Conclusion and Critique}
Two different models for spiral galaxies are proposed here. Both explain the observed rotation curves. However, they are
distinctly different from models based on dark matter and MOND. The strength of these models is in the absence of any postulated
new kinds of matter or forces. However, these models have their weaknesses. Some of them are as follows.
\begin{itemize}
	\item Hyperbolic trajectories of stars are expected. Hence, the non-observation of radial speeds needs to be explained.
	However, it is noticed that significant radial speeds are achieved only by stars at large radial distances. Such stars
	would be difficult to observe because of reduced brightness and population.
	\item For the spinner model, particles within the disk are assumed to have roughly circular orbits due to local forces
	that mimic rigid body forces. Further analysis of the nature of such local forces is deemed necessary.
	\item The Elvis model assumes approximate thermal equilibrium of explosion fragments of all sizes. This is not too strong an
	assumption as all that is needed is the near-light speeds for the smaller plume particles.
\end{itemize}
 


\begin{thebibliography}{99}
\bibitem{rubin0} V.~C.~Rubin and W.~K.~Ford, Astrophysical Journal, {\bf 159}, 379-403, (1970).
\bibitem{rubin1} V.~C.~Rubin, W.~K.~Ford and N.~Thonnard, Astrophysical Journal, {\bf 225}, L107-L111, (1978).
\bibitem{rubin2} V.~C.~Rubin, W.~K.~Ford and N.~Thonnard, Astrophysical Journal, {\bf 238}, 471-487, (1980).
\bibitem{bosma} A.~Bosma, Astronomical Journal, {\bf 86}, 1791-1824, (1981).
\bibitem{persic0} M.~Persic and P.~Salucci, Mon.~Not.~R.~Astron.~Soc. {\bf 245}, 577-581, (1990).
\bibitem{persic1} M.~Persic, P.~Salucci and F.~Stel, Mon.~Not.~R.~Astron.~Soc. {\bf 281}, 27-47, (1996).
\bibitem{corbelli} E.~Corbelli and P.~Salucci, Mon.~Not.~R.~Astron.~Soc. {\bf 311}, 441-447, (2000).
\bibitem{gentile} G.~Gentile, P.~Salucci, U.~Klein, D.~Vergani and P.~Kalberla, Mon.~Not.~R.~Astron.~Soc. {\bf 351}, 903-922, (2004).
\bibitem{merritt} D.~Merritt, J.~F.~Navarro, A.~Ludlow, and A.~Jenkins, Astrophysical Journal, {\bf 624}, L85-L88, (2005).
\bibitem{yegorova} I.~A.~Yegorova and P.~Salucci, Mon.~Not.~R.~Astron.~Soc. {\bf 377}, 507-515, (2007).
\bibitem{duffy} A.~R.~Duffy, J.~Schaye, S.~T.~Kay, C.~D.~Vecchia, R.~A.~Battye and C.~M.~Booth, Mon.~Not.~R.~Astron.~Soc. {\bf 405}, 2161-2178, (2010).
\bibitem{mcgaugh0} S.~S.~McGaugh, and W.~J.~G.~de~Blok, Astrophysical Journal, {\bf 499}, 66-81, (1998).
\bibitem{mcgaugh1} S.~S.~McGaugh, Astrophysical Journal Letters, {\bf 832}, L8, (2016).
\bibitem{mcgaugh2} S.~S.~McGaugh, F.~Lelli and J.~M.~Schombert, Phys.~Rev.~Lett. {\bf 117}, 201101, (2016).
\end{thebibliography}
\end{document}